\documentclass[english]{article}
\usepackage[T1]{fontenc}
\usepackage[latin1]{inputenc}
\usepackage{geometry}
\usepackage{amsmath}
\usepackage{setspace}
\usepackage{amssymb}

\makeatletter
\newcommand{\lyxaddress}[1]{
\par {\raggedright #1
\vspace{1.4em}
\noindent\par}
}

\usepackage{babel}
\makeatother
\begin{document}

\title{Quaternion-Octonion Analyticity for Abelian and Non-Abelian Gauge
Theories of Dyons }

\author{P. S. Bisht$^{\text{(1)}}$ and O. P. S. Negi$^{\text{(}2)}$%
\thanks{Permanent Address- Department of Physics, Kumaun University, S. S.
J. Campus, Almora -263601 (Uttarakhand) INDIA%
}}

\maketitle

\lyxaddress{\begin{center}
$^{\text{(1)}}$ Department of Physics\\
Kumaun University\\
S. S. J. Campus\\
Almora - 263601 (Uttarakhand) INDIA
\par\end{center}}

\lyxaddress{\begin{center}
$^{\text{(2)}}$ Institute of Theoretical Physics \\
Chinese Academy of Sciences\\
 KITP Building Room No.- 6304\\
Hai Dian Qu Zhong Guan Chun Dong Lu \\
55 Hao , Beijing - 100080, P. R. CHINA
\par\end{center}}

\lyxaddress{\begin{center}
Email: - ps\_bisht123@rediffmail.com\\
ops\_negi@yahoo.co.in 
\par\end{center}}

\begin{abstract}
Einstein- Schrödinger (ES) non-symmetric theory has been extended
to accommodate the Abelian and non-Abelian gauge theories of dyons
in terms of the quaternion-octonion metric realization. Corresponding
covariant derivatives for complex, quaternion and octonion spaces
in internal gauge groups are shown to describe the consistent field
equations and generalized Dirac equation of dyons. It is also shown
that quaternion and octonion representations extend the so-called
unified theory of gravitation and electromagnetism to the Yang-Mill's
fields leading to two $SU(2)$ gauge theories of internal spaces due
to the presence of electric and magnetic charges on dyons. 


\end{abstract}

\section{Introduction}

Einstein- Schrödinger (ES) theory \cite{key-1,key-2}, a generalization
of general relativity, allows a non-symmetric fundamental tensor and
connection. It contains a non symmetric metric whose real symmetric
part is described as general relativity while imaginary ( a skew symmetric)
part was taken by Einstein \cite{key-1} as proportional to the electromagnetic
tensor. Research in this direction ultimately proved fruitless; the
desired classical unified field theory was not found and the interpretation
of the skew symmetric part of the metric, as an electromagnetic field
tensor, has been shown physically incorrect \cite{key-3}. However,
Moffat \cite{key-4} and others \cite{key-5} showed that instead
of electromagnetism the anti-symmetric part of the generalized metric
tensor represents a kind of non-symmetric gravitational field which
is free from ghost poles, tachyons and higher-order poles, and there
are no problems with asymptotic boundary conditions. Einstein-Schrödinger
(ES) theory has also been modified and extended \cite{key-6} to include
a large cosmological constant (caused by zero-point fluctuations)
and sources (spin-$0$ and spin- $\frac{1}{2}$ ). In the weak field
approximation where interaction between fields is not taken into account,
the resulting theory is characterized by a symmetric rank-2 tensor
field (gravity), an anti-symmetric tensor field, and a constant characterizing
the mass of the antisymmetric tensor field. The anti-symmetric tensor
field is found to satisfy the equations of a Maxwell-Proca massive
anti-symmetric tensor field. Furthermore, the theory permits one or
more \char`\"{}running constants\char`\"{}: it allows the mass of
the anti-symmetric field, the coupling constant between the anti-symmetric
field and matter, and the gravitational constant to vary as functions
of space and time coordinates. In other words, non-symmetric gravitation
theory (NGT) can be described as a theory that involves a symmetric
tensor field (gravity), an anti-symmetric tensor field, and one or
more scalar fields. On the other hand, Borchsenius \cite{key-7} developed
a principle of correspondence and constructed an unified non-symmetric
theory which includes gravity, electromagnetism and Yang Mills field
theory. Unfortunately, none of the non-symmetric unified models survived
as a plausible theory. Besides the problem of spin-$0$and not the
spin- $1$content of the antisymmetric part of the metric, it was
shown by Damour et al \cite{key-8} that the Einstein theory and those
which are based on the Einstein Lagrangian, exhibit negative-energy
(ghosts) radiative modes and accordingly the Borchsenius \cite{key-7}
theory , which includes the Yang Mills field in Bonor- Moffat- Boal
(BMB) \cite{key-5} theory, has the same problems. However, the inconsistencies
and cure as well as problems and hopes have always been challenging
issues \cite{key-8,key-9} in non-symmetric gravity theories and still
the status of ES or NGT is not clear. Moreover, Morques and Oliveira
\cite{key-10} has developed the quaternion-octonion geometrical extension
and interpretation of Einstein- Schrödinger (ES) non-symmetric theory
which includes consistently \cite{key-11} the Bonor- Moffat- Boal
(BMB) \cite{key-5} and Borchsenius \cite{key-7} theories. It is
shown \cite{key-10,key-11} that the real algebra describes general
theory of relativity, the complex algebra gives the interpretation
of Einstein- Schrödinger (ES) non-symmetric theory and Borchsenius
theory \cite{key-7} is interpreted in terms of quaternions isomorphic
to $SU(2)$ group. Similar work has been done by Ragusa \cite{key-12}
by enlarging NGT to include the Yang-Mills field theory and it is
shown that the anti-symmetric part of the metric tensor ( $2×2$ matrix),
describes both types of field equations namely the electromagnetism
and Yang-Mills field in the flat space linear approximation. Yet,
the inconsistencies and cure as well as problems and hopes have always
been challenging issues \cite{key-12} in non-symmetric gravity theories
. On the other hand, despite of the potential importance of monopoles\cite{key-13}
and dyons \cite{key-14}, the formalism necessary to describe them
has been clumsy and not manifestly covariant \cite{key-15,key-16}.
So, a self consistent and manifestly covariant theory of generalized
electromagnetic fields of dyons (particle carrying electric and magnetic
charges) has been constructed \cite{key-17,key-18,key-19,key-20}
in terms of two four - potentials \cite{key-21} to avoid the use
of controversial string variables. The generalized charge, generalized
four - potential, generalized field tensor, generalized field vector
and generalized four - current density associated with dyons have
been described as complex quantities with their real and imaginary
parts as electric and magnetic constituents.

So, without going in to the controversies of ES or NGT theories, and
keeping in mind the potential importance of monopoles (or dyons),
in the present paper, we have extended the quaternion-octonion generalization
of non-symmetric metric theory developed by Morques et al \cite{key-10,key-11}
to accommodate the Abelian and non-Abelian gauge theories of the generalized
fields of dyons (particles carrying simultaneous electric and magnetic
charges). We have applied here the non-symmetric metric theory and
the corresponding affine geometry for three different spaces over
the field of complex, quaternion and octonion hyper complex number
systems. It has also been shown that the symmetric part of the unified
metric theory is associated with gravity while anti-symmetric part
is described as the generalized electromagnetic field tensor of dyons.
Extension of the metric by quaternionic tangent space has been shown
to describe the total curvature in terms of gravitation and electromagnetism,
along with the non-Abelian Yang-Mill's field ( internal quaternion
curvature) while the further extension of the metric theory to the
case of octonions leads the internal octonionic curvature which gives
rise to two different Yang-Mill's gauge fields. So, the present theory
describes the combined gauge structures $GL(R)\otimes U(1)_{e}\otimes U(1)_{m}\otimes SU(2)_{e}\otimes SU(2)_{m}$
where $GL(R)$ describes Gravity, $U(1)_{e}$demonstrates the electromagnetism
due to the presence of electric charge, $U(1)_{m}$ is responsible
for the electromagnetism due to magnetic monopole, $SU(2)_{e}$demonstrates
the Yang Mill's field due to the presence of electric charge while
$SU(2)_{m}$gives rise the another Yang-Mills field due to the presence
of magnetic monopole. It has also been shown that this unified picture
reproduces the Gravity, electromagnetism and theory of Yang-Mill's
field in the absence of magnetic monopole. Accordingly we have obtained
the generalized Dirac equation for dyons from the covariant derivatives
in terms of complex, quaternionic and octonionic tangent spaces.

\section{Quaternion-Octonion Generalization of Non-Symmetric Metric }

The real formulation of non-symmetric theory is expressed \cite{key-10,key-11}
in terms of the real tensor $g_{\mu\nu}$ as,

\begin{eqnarray}
g_{\mu\nu} & = & h_{\mu\nu}+k_{\mu\nu}\label{eq:1}\end{eqnarray}
and its conjugate is accordingly given by

\begin{eqnarray}
\overline{g_{\mu\nu}} & = & h_{\mu\nu}-k_{\mu\nu}=h_{\mu\nu}+k_{\mu\nu}=g_{\nu\mu}\label{eq:2}\end{eqnarray}
 So, in Non-Riemannian space - time, ES non - symmetric theory is
described in terms of an $n-$ dimensional internal space and thus
the line element on the curved space-time is written \cite{key-10,key-11}
as

\begin{eqnarray}
ds^{2} & = & \frac{1}{n}Tr(G_{\mu\nu}dx^{\mu}dx^{\nu})\label{eq:3}\end{eqnarray}
where

\begin{eqnarray}
G_{\mu\nu} & = & (G_{\mu\nu b}^{a}(x)),\qquad\forall a,b=1,2...n.\label{eq:4}\end{eqnarray}
is a matrix of internal space such that

\begin{eqnarray}
(\frac{1}{n})TrG_{\mu\nu} & = & g_{\mu\nu}.\label{eq:5}\end{eqnarray}
Here $g_{\mu\nu}$ is the metric of the ES asymmetric theory and $Tr$
is acting on internal matrices. We have

\begin{eqnarray}
G_{\mu\nu}^{\dagger} & = & G_{\nu\mu}\label{eq:6}\end{eqnarray}
where the $(\dagger)$ operation is used for the Hermitian conjugate.
Here $G_{\mu\nu}$ is an object with two matrix indexes $a$ and $b$
in internal space supposedly restricted to the internal space of $2\times2$
unitary matrices of $SU(2)$ symmetry group. So, each object in this
space may then be written as a linear combination of four linearly
independent matrices $\tau_{\mu}$ , $(\mu=0,1,2,3)$, where $\tau_{0}=0$
and $\tau_{j}^{\dagger}=\tau_{j}\quad(j=1,2,3)$. Hence the metric
(\ref{eq:4}) may now be written as

\begin{eqnarray}
G_{\mu\nu} & = & g_{\mu\nu}\tau_{0}+f_{\mu\nu i}\tau_{i}\label{eq:7}\end{eqnarray}
with

\begin{eqnarray}
g_{\mu\nu} & = & g_{\mu\nu}+iF_{\mu\nu},\label{eq:8}\end{eqnarray}
where $F_{\mu\nu}$ is the Maxwell's tensor , $f_{\mu\nu i}$ represents
the Yang-Mills field strength and for brevity, we have set all the
coefficients (or constants) as unity. The internal covariant derivative
of a vector (or a spinor in spin space) $\psi^{a}=\psi^{a}(x),\;\forall\, a=1,2$,
is defined as 

\begin{eqnarray}
\psi_{||\mu}^{a} & = & \psi_{,\mu}^{a}+\Gamma_{\mu b}^{a}\psi^{b}.\label{eq:9}\end{eqnarray}
Here the affinity $\Gamma_{\mu}=(\Gamma_{\mu b}^{a}(x))$ is the object
which makes $\psi_{,\mu}^{a}$ to transform like a vector under transformations
in the internal space. $\Gamma_{\mu}$can be related with the gauge
potential as 

\begin{eqnarray}
\Gamma_{\mu} & = & iC_{\mu}.\tau\label{eq:10}\end{eqnarray}
and obeys the following internal transformations law as

\begin{eqnarray}
\Gamma_{\mu}^{'} & = & U(x)\Gamma_{\mu}U^{-1}(x)-\frac{\partial U(x)}{\partial x^{\mu}}U^{-1}(x)\label{eq:11}\end{eqnarray}
where $U(x)$ is defined in terms of the internal transformation matrices
of local $SU(2)$ gauge group. In the curved space-time $\Gamma_{\mu}$
, transforms like a vector. Thus the internal curvature is defined
as 

\begin{eqnarray}
\psi_{||\mu\nu}^{a}-\psi_{||\nu\mu}^{a} & = & P_{\mu\nu b}^{a}\psi^{b}.\label{eq:12}\end{eqnarray}
Here $P_{\mu\nu b}^{a}$ is the curvature in the internal space, i.e.

\begin{eqnarray}
P_{\mu\nu} & = & \Gamma_{\mu,\nu}-\Gamma_{\nu,\mu}-[\Gamma_{\mu},\Gamma_{\nu}]\label{eq:13}\end{eqnarray}

\begin{eqnarray}
P_{\mu\nu} & = & -P_{\nu\mu}\label{eq:14}\end{eqnarray}
In case of complex tangent space, the $U(x)$ ''internal transformation
matrix'' is described by the matrices of the internal gauge group
$U(1)$. Hence the transformation laws of an object in the complex
$C$ - space, $K$ may be written as,

\begin{eqnarray}
K' & = & U(1)K\label{eq:15}\end{eqnarray}
where $U(1)$ stands for a unitary $1\times1$ (local) transformation
matrix, $U(1)=e^{i\phi(x)}$, and

\begin{eqnarray}
\overline{K}' & = & \overline{U}(1)\overline{K}\label{eq:16}\end{eqnarray}
where $\overline{U}(1)=U^{-1}(1)=e^{-i\phi(x)}$ . Accordingly, the
{}``internal connection'' $C_{\nu}$ is transformed as \begin{eqnarray}
C\,'_{\nu} & = & U(1)C_{\nu}U^{-1}(1)-U_{,\nu}(1)U^{-1}(1)\label{eq:17}\end{eqnarray}
It should be noted that the connection $C_{\nu}$ transforms as a
vector under space - time transformations. In a particular case where
the internal transformations are represented by the matrices $U(1)=1+i\phi$,
the connection $C_{\nu}$ transforms in first order as,

\begin{eqnarray}
C_{\nu}' & = & C_{\nu}+i\phi_{,\nu}\label{eq:18}\end{eqnarray}
which follows the gauge transformation law for an electromagnetic
potential. 

In the Borchsenius theory \cite{key-7} , the vector space is described
in terms of Pauli matrices which can be reinterpreted as quaternions
\cite{key-10,key-11} to describe quaternion tangent space. In this
case, we take $e_{j}=i^{-1}\sigma_{j},\,\,\forall\, j=1,2,3;$ $i=\sqrt{-1}$;
$\sigma_{i}$ are Pauli matrices and $e_{j}$ are quaternion basis
elements satisfying the following multiplication relation

\begin{eqnarray}
e_{j}\, e_{k} & = & -\delta_{jk}e_{0}+\varepsilon_{jkl}e_{k}\label{eq:19}\end{eqnarray}
where $e_{0}$ is the unit element of the quaternion algebra i.e.
$e_{0}=\sigma_{0}$ and $\delta_{jk}$is the Kronecker delta symbol.
So, the metric in quaternionic space - time undergoes with the symmetry
property given by equation (\ref{eq:6}) where the Hermitian conjugation
operation is carried out in terms of the quaternionic internal space
or $Q-$ space and $\Gamma_{\nu}=-C_{\nu}^{a}e_{a}$ is the affinity
in the quaternionic internal space and transforms under the transformation
laws given by equation (\ref{eq:11}). Since, quaternion basis elements
are isomorphic to the algebra of  Pauli spin matrices, we may obtain
other results given by equations (\ref{eq:12}, \ref{eq:13}) and
(\ref{eq:14}) in quaternion tangent space. 

For octonion tangent space, we use the split octonion $O$ algebras
where an octonion $P$ is written \cite{key-22,key-23,key-24} in
the split $O$ algebra as,

\begin{eqnarray}
P & = & au_{0}^{\star}+bu_{0}-n_{k}u_{k}^{\star}+m_{k}u_{k}\qquad\forall\, k=1,2,3\label{eq:20}\end{eqnarray}
where $u_{0}^{\star},u_{0},u_{k},u_{k}^{\star}\,\,\,(\forall k=1,2,3)$
are the the split $O$ basis elements \cite{key-10,key-11,key-22,key-23,key-24}
defined as

\begin{eqnarray}
u_{0}=\frac{1}{2} & (e_{0}+ie_{7}); & \quad u_{0}^{\star}=\frac{1}{2}(e_{0}-ie_{7});\nonumber \\
u_{k}=\frac{1}{2} & (e_{k}+ie_{k+3}); & \quad u_{k}^{\star}=\frac{1}{2}(e_{k}-ie_{k+3}).\label{eq:21}\end{eqnarray}
Here the set of octets $e_{0},$$e_{1},e_{2},e_{3},e_{4},e_{5},e_{6},e_{7}$are
known as the octonion units satisfying the following multiplication
rule

\begin{eqnarray}
e_{0}^{2}=e_{0}=1;\, & e_{0}e_{A}=e_{A}e_{0} & =e_{A};\, e_{A}e_{B}=-\delta_{AB}e_{0}+f_{ABC}e_{C}\,(\forall A,B,C=1,2....6,7);\label{eq:22}\end{eqnarray}
where the structure constants $f_{ABC}$is completely antisymmetric
and takes the value $1$ for following combinations

\begin{eqnarray}
f_{ABC} & =1 & \forall(ABC)=(123);(471);(257);(165);(624);(543);(736).\label{eq:23}\end{eqnarray}
So, an split octonion $P$ given by equation (\ref{eq:20}) is now
be written \cite{key-10,key-11,key-22,key-23,key-24}, in terms of
$2\times2$ Zorn's vector matrix realizations as .\begin{eqnarray}
P\cong Z(P) & = & \left(\begin{array}{cc}
a & -\overrightarrow{n}\\
\overrightarrow{m} & b\end{array}\right).\label{eq:24}\end{eqnarray}
We may also express split octonion algebra in terms of Pauli matrices
which are related with the quaternion basis elements given by equation
(\ref{eq:19}). So, we define the following $2\times2$ Zorn's vector
matrix realizations of split octonion basis elements $u_{0}^{\star},u_{0},u_{k},u_{k}^{\star}\,\,\,(\forall k=1,2,3)$
i.e.

\begin{eqnarray}
Z(u_{0}^{\star}) & = & \left(\begin{array}{cc}
1.e_{0} & 0_{2}\\
0_{2} & 0_{2}\end{array}\right);\qquad Z(u_{0})=\left(\begin{array}{cc}
0_{2} & 0_{2}\\
0_{2} & 1.e_{0}\end{array}\right),\nonumber \\
Z(u_{k}^{\star}) & = & \left(\begin{array}{cc}
0_{2} & -1.e_{k}\\
0_{2} & 0_{2}\end{array}\right);\qquad Z(u_{k})=\left(\begin{array}{cc}
0_{2} & 0_{2}\\
1.e_{k} & 0_{2}\end{array}\right).\label{eq:25}\end{eqnarray}
The octonion conjugation of $P$ is now defined as

\begin{eqnarray}
\overline{P} & = & bu_{0}^{\star}+au_{0}+n_{k}u_{k}^{\star}-m_{k}u_{k}^{\star}\,\,\,\,\,(\forall k=1,2,3)\label{eq:26}\end{eqnarray}
and a Hermitian conjugate of $P$ is expressed as

\begin{eqnarray}
P^{\dagger}=(\overline{P})^{\star} & = & b^{\star}u_{0}^{\star}+a^{\star}u_{0}+n_{k}^{\star}u_{k}^{\star}-m_{k}^{\star}u_{k}\,\,\,(\forall k=1,2,3).\label{eq:27}\end{eqnarray}
As such , we may reformulate equation (\ref{eq:4}) as the $O$ {}``metric''
with its split form \cite{key-10,key-11} as

\begin{eqnarray}
G_{\mu\nu}(x) & = & \left(\begin{array}{cc}
s_{\mu\nu}^{0}e_{0} & -s_{\mu\nu}^{k}e_{\kappa}\\
r_{\mu\nu}^{k}e_{\kappa} & r_{\mu\nu}^{0}e_{0}\end{array}\right)=G_{\mu\nu}(s,r).\label{eq:28}\end{eqnarray}
Here $r_{\mu\nu}^{0}=s_{\mu\nu}^{0}=g_{(\mu\nu)}+iF_{[\mu\nu]};\;\, g_{(\mu\nu)}$
is identified as the symmetric metric (gravity-expressed in terms
of algebra of real numbers$GL(R)$) and $F_{[\mu\nu]}$ is the Maxwell
$U(1)$ valued electromagnetic field strength, while $r_{\mu\nu}^{k}$and
$s_{\mu\nu}^{k}$ are $SU(2)$ valued field strengths of two Yang-Mills
( non-Abelian gauge) fields. So, we get the following symmetry property

\begin{eqnarray}
G_{\mu\nu}^{\dagger}(s,r) & = & G_{\nu\mu}(s,r).\label{eq:29}\end{eqnarray}
 and 

\begin{eqnarray}
G_{\mu\alpha}(s,r)G^{\mu\nu}(s,r) & = & G^{\nu\mu}(s,r)G_{\alpha\mu}(s,r)=\delta_{\alpha}^{\nu}(u_{0}+u_{0}^{*}).\label{eq:30}\end{eqnarray}
Here we agree with the statement of Castro \cite{key-24} that the
most salient feature of the split octonion metric $G_{\mu\nu}$given
by equation (\ref{eq:28}) is that it includes the ordinary space
time metric $g_{\mu\nu}$, in addition to electromagnetism and Yang
Mills fields. Hence it automatically justifies the Kaluza-Klein theory
without introducing extra space-time dimensions. The line element
in the $O$ space - time is thus defined by

\begin{eqnarray}
ds^{2} & = & \frac{1}{4}Tr(dx^{\mu}dx^{\nu}G_{\mu\nu})\label{eq:31}\end{eqnarray}
 while the affinity $\Gamma_{\nu}$ given by equation (\ref{eq:9})
is expressed in the internal octonionic space as

\begin{eqnarray}
\Gamma_{\nu} & = & \left(\begin{array}{cc}
0_{2} & -L_{\nu}.e\\
K_{\nu}.e & 0_{2}\end{array}\right)\label{eq:32}\end{eqnarray}
where $\left\{ L_{\nu}\right\} $ and $\left\{ K_{\nu}\right\} $
are two real four- potentials (gauge connections) analogous to $\left\{ C_{\mu}\right\} $
given by equation (\ref{eq:10}) and discussed above for the quaternionic
case. Like equation (\ref{eq:13}), the octonion curvature $S_{\nu\gamma}$
may then be written as

\begin{eqnarray}
S_{\nu\gamma c}^{a} & = & S_{\nu\gamma c}^{a}(u_{0}+u_{0}^{*})+\delta_{c}^{a}P_{\nu\gamma}\label{eq:33}\end{eqnarray}

\section{Non-Symmetric Metric and Dyonic Fields}

Let us extend the Einstein-Schrödinger non -symmetrical metric in
terms of three different tangent spaces namely complex, quaternionic
and octonionic cases associated with the generalized fields of dyons.

\subsection{Complex Case}

In the complex tangent space case, the non-symmetric metric $g_{\mu\nu}$
given by equation (\ref{eq:1}) is rewritten as

\begin{eqnarray}
g_{\mu\nu} & = & g_{\mu\nu}+ik_{\mu\nu}.\label{eq:34}\end{eqnarray}
where we represent $F_{[\mu\nu]}=k_{\mu\nu}$ as the anti - symmetric
tensor comprising electromagnetic field associated with dyons in the
following manner

\begin{eqnarray}
k_{\mu\nu} & \rightarrow & (F_{\mu\nu}+iF_{\mu\nu}^{d})\label{eq:35}\end{eqnarray}
where $\left\{ F_{\mu\nu}\right\} $ and $\left\{ F_{\mu\nu}^{d}\right\} $
are described as generalized electromagnetic and dual electromagnetic
fields of dyons \cite{key-17,key-18,key-22}. So this extension accommodates
both types of non-symmetric metrics real as well as complex where
one of the Maxwell field is always real. Thus for the dyonic case,
we may identify the internal connection (i.e. $C_{\nu}$) as the generalized
electromagnetic potential $\left\{ V_{\mu}\right\} $ of dyons \cite{key-17,key-18,key-22}
described as ;

\begin{eqnarray}
\left\{ V_{\mu}\right\}  & = & \left\{ A_{\mu}\right\} -i\,\left\{ B_{\mu}\right\} .\label{eq:36}\end{eqnarray}
Hence, in complex case , non - symmetric metric (\ref{eq:35}) is
associated with our generalized electromagnetic field tensor of dyons
\cite{key-17,key-18,key-22} (i.e. the matrix of the internal space
for dyonic fields) as 

\begin{eqnarray}
G_{\mu\nu} & = & F_{\mu\nu}-iF_{\mu\nu}^{d},\nonumber \\
G_{\mu\nu}^{\star} & = & F_{\mu\nu}+iF_{\mu\nu}^{d}\label{eq:37}\end{eqnarray}
where $(\star)$ denotes the complex conjugation. Hence, replacing
the gauge connection $\left\{ C_{v}\right\} $ by our generalized
four potential $\left\{ V_{\nu}\right\} $, we get,

\begin{eqnarray}
G_{\mu\nu} & = & V_{\mu,\nu}-V_{\nu,\mu},\nonumber \\
G_{\mu\nu}^{\star} & = & V_{\mu,\nu}^{\star}-V_{\nu,\mu}^{\star}.\label{eq:38}\end{eqnarray}
As such, correspondingly, we have the following field equations ,

\begin{eqnarray}
G_{\mu\nu,\nu} & = & J_{\mu},\nonumber \\
G_{\mu\nu,\nu}^{\star} & = & J_{\mu}^{\star}\label{eq:39}\end{eqnarray}
where $\left\{ J_{\mu}\right\} $ represents the generalized current
for the dyonic fields given by $\left\{ J_{\mu}\right\} =\left\{ j_{\mu}^{e}\right\} -i\,\left\{ j_{\mu}^{m}\right\} $
with $\left\{ j_{\mu}^{e}\right\} $ and $\left\{ j_{\mu}^{m}\right\} $
are described as the four currents respectively associated with electric
and magnetic charges. Equation (\ref{eq:37}) gives the following
decompositions of electric and magnetic field strengths of dyons i.e.
\begin{eqnarray}
F_{\mu\nu} & = & \frac{1}{2}(G_{\mu\nu}+G_{\mu\nu}^{\star}),\nonumber \\
F_{\mu\nu}^{d} & =- & \frac{1}{2i}(G_{\mu\nu}-G_{\mu\nu}^{\star}).\label{eq:40}\end{eqnarray}
Thus we obtain the following decoupled Generalized Dirac-Maxwell's
(GDM) equations of dyons in terms of electric and magnetic four currents
as 

\begin{eqnarray}
F_{\mu\nu,\nu} & = & \frac{1}{2}(G_{\mu\nu,\nu}+G_{\mu\nu,\nu}^{\star})=\frac{1}{2}(J_{\mu}+J_{\mu}^{\star})=j_{\mu}^{e},\nonumber \\
F_{\mu\nu,\nu}^{d} & =- & \frac{1}{2i}(G_{\mu\nu,\nu}-G_{\mu\nu,\nu}^{\star})=\frac{1}{2}(J_{\mu}-J_{\mu}^{\star})=j_{\mu}^{m}\label{eq:41}\end{eqnarray}
Here dyons are considered as the point particle carrying simultaneous
existence of electric and magnetic charges in terms of two Abelian
$U(1)$ gauge structures. Now replacing internal connection $\left\{ C_{\mu}\right\} $
by generalized potential $\left\{ V_{\mu}\right\} $, we may apply
the following internal transformation law (\ref{eq:22},\ref{eq:23})
as

\begin{eqnarray}
V_{\mu} & \rightarrow & \Omega V_{\mu}^{\star}\Omega^{-1}-(\partial_{\mu}\Omega)\Omega^{-1}\label{eq:42}\end{eqnarray}
and the corresponding equation (\ref{eq:18}) for the dyonic field
is then expressed as

\begin{eqnarray}
V_{\mu} & \rightarrow & V_{\mu}+i\phi_{,\mu}\label{eq:43}\end{eqnarray}
which is the gauge transformation for the generalized electromagnetic
potential of dyons. Similarly we may write

\begin{eqnarray}
V_{\mu}^{\star} & \rightarrow & \Omega V_{\mu}^{\star}\Omega^{-1}-(\partial_{\mu}\Omega)\Omega^{-1}\nonumber \\
\rightarrow & V_{\mu}^{\star} & +i\phi_{,\mu}^{\star}\label{eq:44}\end{eqnarray}
and consequently we get the following decoupled electric and magnetic
$U(1)$ gauge connections 

\begin{eqnarray}
A_{\mu} & = & \frac{1}{2}(V_{\mu}+V_{\mu}^{\star}),\nonumber \\
B_{\mu} & = & \frac{i}{2}(V_{\mu}-V_{\mu}^{\star}).\label{eq:45}\end{eqnarray}
Here $A_{\mu}$ and $B_{\mu}$ represent the electric and magnetic
four - potential of dyonic fields and are the out comes of the non-symmetric
metric in the complex tangent space. As such we may write the covariant
derivative $D_{\mu}$ as

\begin{eqnarray}
D_{\mu} & \rightarrow & \partial_{\mu}+iV_{\mu},\nonumber \\
D_{\nu} & \rightarrow & \partial_{\nu}+iV_{\nu}\label{eq:46}\end{eqnarray}
and therefore

\begin{eqnarray}
[D_{\mu},D_{\nu}] & = & D_{\mu}D_{\nu}-D_{\nu}D_{\mu}=G_{\mu\nu}\label{eq:47}\end{eqnarray}
which satisfies the generalized Maxwell's - Dirac equation for dyonic
fields given equation \ref{eq:39}. As such, without disturbing the
real part of the non symmetric ES metric ( taking it as gravity )
we have successfully extended its imaginary part corresponding to
the generalized fields of dyons in order to reformulate the self-consistent
and manifestly covariant theory of dyons.

\subsection{Quaternion Case}

In order to develop unified quaternionic non - symmetric metric theory
, we use the bi -quaternionic formulation of dyons described earlier
( Shalini Bisht et al \cite{key-20} ) instead of using the metric
of the real quaternionic tangent space since bi -quaternions work
over the filed of complex numbers like ordinary quaternions do with
real numbers. So, the metric given equations (\ref{eq:7},\ref{eq:8})
is now written as

\begin{eqnarray}
G_{\mu\nu} & = & G_{\mu\nu}^{0}e_{0}+G_{\mu\nu}^{j}e_{j}\label{eq:48}\end{eqnarray}
and

\begin{eqnarray}
G_{\mu\nu}^{0} & = & g_{\mu\nu}\Rightarrow g_{\mu\nu}+ik_{\mu\nu}\Rightarrow g_{\mu\nu}+i\,(F_{\mu\nu}-iF_{\mu\nu}^{d});\nonumber \\
G_{\mu\nu}^{j} & \Rightarrow & f_{\mu\nu j}=f_{\mu\nu j}^{e}-if_{\mu\nu j}^{m}\label{eq:49}\end{eqnarray}
where superscript ($e$) and ($m$) are used for electric and magnetic
counter parts of dyons. Accordingly, we may use the properties of
quaternion metric, internal covariant derivative, the transformation
law, curvature etc. for the quaternionic space-time given by equations
(\ref{eq:9}) to (\ref{eq:14}). 

Let us now define the the covariant derivative \cite{key-18} in quaternionic
non - symmetric metric theory of dyonic fields as 

\begin{eqnarray}
D_{\mu} & \rightarrow & \partial_{\mu}+V_{\mu}e_{0}+V_{\mu}^{a}e_{a},\nonumber \\
D_{\nu} & \rightarrow & \partial_{\nu}+V_{\nu}e_{0}+V_{\nu}^{a}e_{a}\quad(a=1,2,3)\label{eq:50}\end{eqnarray}
 which gives the complex Abelian and non-Abelian $U(1)\times SU(2)$
gauge structure. The second term in the right hand side of equation
(\ref{eq:50}) represents the electromagnetic $U(1)$ part while the
third term represents the non - Abelian $SU(2)$ part of Yang - Mill's
field spanned in the term of quaternion basis elements. Then we get

\begin{eqnarray}
[D_{\mu},D_{\nu}]\psi & = & (D_{\mu}D_{\nu}-D_{\nu}D_{\mu})\psi=(G_{\mu\nu}e_{0}+G_{\mu\nu}^{a}e_{a})\psi\label{eq:51}\end{eqnarray}
which describes $U(1)\times SU(2)$ gauge field strengths for generalized
fields of dyons. In equation (\ref{eq:51}) we have

\begin{eqnarray}
G_{\mu\nu} & = & \partial_{\mu}V_{\nu}-\partial_{\nu}V_{\mu},\qquad G_{\mu\nu}^{a}=\partial_{\mu}V_{\nu}^{a}-\partial_{\nu}V_{\mu}^{a}\label{eq:52}\end{eqnarray}
and subsequently we get the following field equations 

\begin{eqnarray}
G_{\mu\nu,\nu} & = & D_{\nu}G_{\mu\nu}=j_{\mu};\qquad G_{\mu\nu,\nu}^{a}=D_{\nu}G_{\mu\nu}^{a}=j_{\mu}^{a};\label{eq:53}\end{eqnarray}
where $j_{\mu}$ and $j_{\mu}^{a}$ are the generalized current corresponding
to the electromagnetic part $U(1)$ and non - Abelian part $SU(2)$
respectively for the dyonic fields. So, we get the following continuity
equation for generalized fields of dyons as

\begin{eqnarray}
\partial^{\mu}J_{\mu} & = & 0,\label{eq:54}\end{eqnarray}
but for non-Abelian gauge fields, we get 

\begin{eqnarray}
\partial^{\mu}J_{\mu}^{a}\neq0;\,\qquad & D^{\mu}J_{\mu} & =0\label{eq:55}\end{eqnarray}
where

\begin{eqnarray}
J_{\mu} & = & J_{\mu}e_{0}+J_{\mu}^{a}e_{a}\label{eq:56}\end{eqnarray}
which is the $U(1)\times SU(2)$ gauge structure of the generalized
current associated with dyons consisting point like $U(1)$ gauge
structure of Abelian four current $\left\{ J_{\mu}\right\} $ followed
by $SU(2)$ like extended Yang - Mill's gauge structure $\left\{ J_{\mu}^{a}\right\} $
as the non - Abelian gauge current.

\subsection{Octonion Case}

Octonionic tangent space has been defined in terms of its split basis.
Its metric is also defined in split form by equations (\ref{eq:28},\ref{eq:29})
while line element in the $O$ space - time is expressed by equations
(\ref{eq:31}) and other properties are given by equations (\ref{eq:32}-\ref{eq:33}).
Octonionic gauge formulation of dyonic fields has also been discussed
by us ( Shalini Dangwal et al \cite{key-22} ). As such, we may straight
forwardly write the covariant derivative for the dyonic fields in
split octonion form as,

\begin{eqnarray}
D_{\mu} & \rightarrow & \left(\begin{array}{cc}
\partial_{\mu}+V_{\mu} & -V_{\mu}^{a}e_{a}\\
V_{\mu}^{a*}e_{a} & \partial_{\mu}+V_{\mu}^{a}\end{array}\right);\qquad D_{\nu}\rightarrow\left(\begin{array}{cc}
\partial_{\nu}+V_{\nu} & -V_{\nu}^{a}e_{a}\\
V_{\nu}^{a*}e_{a} & \partial_{\nu}+V_{\nu}^{a}\end{array}\right).\label{eq:57}\end{eqnarray}
Then we get 

\begin{eqnarray}
[D_{\mu},D_{\nu}] & = & \left(\begin{array}{cc}
F_{\mu\nu} & -\overrightarrow{F_{\mu\nu}^{a}}.\overrightarrow{e_{a}}\\
\overrightarrow{f_{\mu\nu}^{a}}.\overrightarrow{e_{a}} & f_{\mu\nu}\end{array}\right)=G_{\mu\nu}\label{eq:58}\end{eqnarray}
where

\begin{eqnarray}
F_{\mu\nu} & = & \partial_{\mu}V_{\nu}-\partial_{\nu}V_{\mu},\nonumber \\
F_{\mu\nu}^{a} & = & \partial_{\mu}V_{\nu}^{a}-\partial_{\nu}V_{\mu}^{a}+i\varepsilon_{abc}V_{\mu}^{b}V_{\nu}^{c},\nonumber \\
f_{\mu\nu} & = & \partial_{\mu}V_{\nu}^{\star}-\partial_{\nu}V_{\mu}^{\star},\nonumber \\
f_{\mu\nu}^{a} & = & \partial_{\mu}V_{\nu}^{a\star}-\partial_{\nu}V_{\mu}^{a\star}+i\varepsilon_{abc}V_{\mu}^{b\star}V_{\nu}^{c\star}.\label{eq:59}\end{eqnarray}
Therefore we may obtain the following split form of field equation
as 

\begin{eqnarray}
D_{\mu}G_{\mu\nu} & = & \left(\begin{array}{cc}
j_{\nu} & -j_{\nu}^{a}e_{a}\\
k_{\nu}^{a}e_{a} & k_{\nu}\end{array}\right)=J_{\nu}\label{eq:60}\end{eqnarray}
and the $U(1)\times SU(2)$form of generalized continuity equation
as 

\begin{eqnarray}
D_{\nu}J_{\nu} & = & 0.\label{eq:61}\end{eqnarray}
as such, the octonion extension of unified non-symmetric metric for
the case of dyons is described in terms two $U(1)$ Abelian ( electromagnetic)
and two $SU(2)$ non Abelian ( Yang-Mills field). Thus for the case
of quaternions and octonions we need not to define the Yang Mills
field by hand. The difference between bi -quaternion and octonion
formulations is that bi-quaternions are non-commutative but associative
while the octonions are neither commutative nor associative and in
split basis the role of associativity is played by the alternativity.
Octonion has the advantage to work in higher dimensional space time.
We may now discuss the decomposition of theories in terms of electric
and magnetic charges in the following manner.

\subsubsection{(Electric Case)}

In this particular case (electric case) if we put that $V_{\mu}=V_{\mu}^{\star}$
i.e. $A_{\mu}-iB_{\mu}=A_{\mu}+iB_{\mu}$$\Rightarrow$ $B_{\mu}=0$
or giving rise to $V_{\mu}=A_{\mu}$. Hence we get the following split
octonion representation of covariant derivative in the absence of
magnetic monopole i.e. 

\begin{eqnarray}
D_{\mu} & \rightarrow & \left(\begin{array}{cc}
\partial_{\mu}+A_{\mu} & -A_{\mu}^{a}e_{a}\\
A_{\mu}^{a}e_{a} & \partial_{\mu}+A_{\mu}\end{array}\right);\qquad D_{\nu}\rightarrow\left(\begin{array}{cc}
\partial_{\nu}+A_{\nu} & -A_{\nu}^{a}e_{a}\\
A_{\nu}^{a}e_{a} & \partial_{\nu}+A_{\nu}\end{array}\right)\label{eq:62}\end{eqnarray}
and then we get

\begin{eqnarray}
[D_{\mu},D_{\nu}] & = & \left(\begin{array}{cc}
F_{\mu\nu} & -\overrightarrow{F_{\mu\nu}^{a}}.\overrightarrow{e_{a}}\\
\overrightarrow{f_{\mu\nu}^{a}}.\overrightarrow{e_{a}} & f_{\mu\nu}\end{array}\right)=E_{\mu\nu}.\label{eq:63}\end{eqnarray}
Consequently 

\begin{eqnarray}
D_{\mu}E_{\mu\nu} & = & \left(\begin{array}{cc}
j_{\nu} & -j_{\nu}^{a}e_{a}\\
j_{\nu}^{a}e_{a} & j_{\nu}\end{array}\right)=J_{\nu}\label{eq:64}\end{eqnarray}
which is the split octonion form of generalized $U(1)\times SU(2)$
field equation where the diagonal elements represent the Maxwell's
equation while the off diagonal elements describe the Yang Mills gauge
fields in absence of magnetic monopole.

\subsubsection{(Magnetic Case)}

In this particular case (electric case) if we put that $V_{\mu}=-V_{\mu}^{*}$
i.e. $A_{\mu}-iB_{\mu}=-A_{\mu}-iB_{\mu}$$\Rightarrow$ $A_{\mu}=0\Rightarrow$$V_{\mu}=-iB_{\mu}$
and hence $V_{\mu}^{\star}=iB_{\mu}$. Therefore , have

\begin{eqnarray}
D_{\mu} & \rightarrow & \left(\begin{array}{cc}
\partial_{\mu}-iB_{\mu} & iB_{\mu}^{a}e_{a}\\
iB_{\mu}^{a}e_{a} & \partial_{\mu}+iB_{\mu}\end{array}\right),\nonumber \\
D_{\nu} & \rightarrow & \left(\begin{array}{cc}
\partial_{\nu}-iB_{\nu} & iB_{\nu}^{a}e_{a}\\
iB_{\nu}^{a}e_{a} & \partial_{\nu}+iB_{\nu}\end{array}\right).\label{eq:65}\end{eqnarray}
and 

\begin{eqnarray}
[D_{\mu},D_{\nu}] & = & \left(\begin{array}{cc}
F_{\mu\nu} & -\overrightarrow{F_{\mu\nu}^{a}}.\overrightarrow{e_{a}}\\
\overrightarrow{f_{\mu\nu}^{a}}.\overrightarrow{e_{a}} & f_{\mu\nu}\end{array}\right)=H_{\mu\nu}.\label{eq:66}\end{eqnarray}
and

\begin{eqnarray}
D_{\mu}H_{\mu\nu} & = & \left(\begin{array}{cc}
k_{\nu} & -k_{\nu}^{a}e_{a}\\
k_{\nu}^{a}e_{a} & k_{\nu}\end{array}\right)=K_{\nu}\label{eq:67}\end{eqnarray}
which is the split octonionic form of generalized $U(1)\times SU(2)$
field equations where the diagonal elements represent the dual Maxwell
equation i.e for pure magnetic monopole and off diagonal elements
describe the Yang Mills gauge fields in in the absence of electric
charge.

\section{Generalized Dirac Equations for Dyons}

We may now adopt the fore going analysis to obtain the Dirac equation
for dyons on using the ES non - symmetric theory. The simplest free
particle Dirac equation is given by

\begin{eqnarray}
(\gamma^{\mu}\partial_{\mu}+\kappa)\psi & = & 0.\label{eq:68}\end{eqnarray}
and to write the interacting form of Dirac equation one has to replace
the partial derivative $\partial_{\mu}$ by covariant derivative $D_{\mu}$.
So, we follow the same process and write the generalized Dirac equation
for particles carrying electric and magnetic charges (i.e. dyons).
Replacing the partial derivative $\partial_{\mu}$ by covariant derivative
$D_{\mu}$, we may write following form of equation of a Dirac particle
in generalized electromagnetic fields of dyons as 

\begin{eqnarray}
(\gamma^{\mu}D_{\mu}+\kappa)\psi & = & 0\label{eq:69}\end{eqnarray}
where we have used the natural units of $c=\hbar=1$ and $D_{\mu}$is
covariant derivative in complex, quaternion and octonion tangent spaces
of Einstein-Schrödinger non-symmetric theory. For complex case the
covariant derivative is illustrated as

\begin{eqnarray}
D_{\mu} & \rightarrow & \partial_{\mu}+iq^{\star}V_{\mu}\label{eq:70}\end{eqnarray}
where

\begin{eqnarray}
q^{\star}V_{\mu} & \rightarrow & eA_{\mu}+gB_{\mu}.\label{eq:71}\end{eqnarray}
Thus the Dirac equation is

\begin{eqnarray*}
\{\gamma_{\mu}(\partial_{\mu}-ieA_{\mu}-igB_{\mu})+\kappa\}\psi & = & 0\end{eqnarray*}
or

\begin{eqnarray}
\{\gamma_{\mu}D_{\mu}+\kappa\}\psi & = & 0\label{eq:72}\end{eqnarray}
which is invariant under gauge transformation as well. For quaternion
case, we may write the covariant derivative as

\begin{eqnarray}
D_{\mu} & \rightarrow & \partial_{\mu}-i(q^{\star}V_{\mu})e_{0}-(q^{\star a}V_{\mu}^{a})e_{a}\label{eq:73}\end{eqnarray}
where

\begin{eqnarray}
q^{\star}V_{\mu} & \rightarrow & eA_{\mu}+gB_{\mu};\quad q^{\star a}V_{\mu}^{a}\rightarrow\varepsilon A_{\mu}^{a}+\varepsilon'B_{\mu}^{a}\label{eq:74}\end{eqnarray}
where $\varepsilon$ and $\varepsilon'$ are the Yang - Mill's coupling
constants associated with the isotropic coupling parameters of electric
and magnetic charges respectively. Similarly the Dirac equation for
generalized fields of dyons in octonionic tangent space is described
by equation (\ref{eq:69}) where the covariant derivative is given
in the split octonion form as

\begin{eqnarray}
D_{\mu} & \rightarrow & \left(\begin{array}{cc}
\partial_{\mu}+q^{*}V_{\mu} & -qV_{\mu}^{a}e_{a}\\
qV_{\mu}^{*a}e_{a} & \partial_{\mu}+qV_{\mu}^{*}\end{array}\right)\label{eq:75}\end{eqnarray}
which is double fold structure of quaternionic tangent space and is
described in terms of Zorn's vector matrix realization of split octonion
basis elements.

\section{Discussion and Conclusion }

It is note worthy to include here, a motivation for the use of split
octonion algebra instead of real octonion because the split octonion
have the advantages to work in terms of matrix realizations while
the due to non associativity of real octonions, it is impossible to
write their correspondence with the matrix realizations. Secondly,
the real octonion forms a metric in eight dimensional structure while
the split octonion has the two $(4,4)$ (four fold ) degeneracy in
complex space time and has the direct correspondence with the bi-quaternions.
We may also develop a similar theory using the real octonion but in
that case it will hard to give the four dimensional correspondence.
So, this is why the quaternion-octonions play an important role in
order to understand the physical theories of higher dimensional supersymmetry
and super gravity etc. As we have mentioned that the octonions consist
seven imaginary units resulting to seven permutations of $SU(2)$
Yang - Mill's fields. So we have the scope to enlarge the metric without
putting the additional structure of space -time by hand and accordingly
there is a possibility to define covariant derivative of a vector
obtained in terms of octonionic vector potential and the octonion
curvature. The automorphism group of octonion algebra is the $14-$
dimensional $G_{2}$ group \cite{key-23,key-24} which admits a $SU(3)$
sub-group and leaves the idempotent $u_{0}$and $u_{0}^{\star}$of
split octonion algebra as invariant. We have established the connection
between real and split basis of octonions and accordingly developed
our present formulation. Due to the lack of associativity in octonion
representation we have described octonion basis elements in terms
of Zorn's vector matrix realizations where octonions are represented
as the double fold degeneracy in terms of quaternion variables to
maintain the consistency in our theory of dyonic fields. Equation
(\ref{eq:34}) represents the non - symmetric metric in the complex
tangent space for dyonic fields, where $k_{\mu\nu}$ is the anti -
symmetric tensor associated with the generalized fields of dyons.
Because the antisymmetric part has been described as further complex
quantity our theory removes the conflicts that Maxwell tensor is real
or imaginary and leaves all other good points of ES or NGT metric
untouched. The Dyon field tensor is expressed by equations (\ref{eq:35})
and (\ref{eq:36}) in terms of electromagnetic field strengths associated
with electric and magnetic sources. In this theory we have replaced
the internal transformation $C_{\nu}$ by generalized gauge potential
$V_{\nu}$ of dyons. It has been shown that the anti - symmetric part
of the metric leads to generalized field equations of dyons discussed
by equation (\ref{eq:39}). Accordingly, we have obtained the electric
and magnetic field tensors from the generalized one for dyons as discussed
by equation (\ref{eq:40}). Consequently, equation (\ref{eq:41})
describes the electric and magnetic four - currents obtained from
the corresponding field tensors of dyons which are considered as the
particles carrying simultaneous existence of electric and magnetic
charges. Equations (\ref{eq:42}) to (\ref{eq:44}) are the unitary
internal transformations for the dyonic gauge potential. Equation
(\ref{eq:46}) represents the covariant derivative in the complex
tangent space for dyonic fields, with the help of which we have obtained
equations (\ref{eq:47}) and (\ref{eq:48}), which in fact represent
the differential forms of generalized Maxwell's - Dirac equation for
dyonic fields. Equation (\ref{eq:49}) expresses the covariant derivative
for dyonic fields in the quaternionic tangent space of the non-symmetric
theory in which the second term represents the electromagnetic part
while the third term represents the non - Abelian part of Yang - Mill's
field in terms of quaternion basis vectors. Also with the help of
equation (\ref{eq:49}) we have obtained equation (\ref{eq:50}),
which describes $U(1)\times SU(2)$ gauge structure of generalized
quaternion tangent space. In equation (\ref{eq:50}) $G_{\mu\nu}$
and $G_{\mu\nu}^{a}$ are the gauge field strengths of Abelian and
non - Abelian fields of dyons. In equation (\ref{eq:52}) $J_{\mu}$
and $J_{\mu}^{a}$ are generalized currents corresponding to the electromagnetic
$U(1)$ part and non-Abelian $SU(2)$ part respectively for dyonic
fields. Equations (\ref{eq:53}) and (\ref{eq:54}) represent the
continuity equation where $J_{\mu}$ is expressed by equation (\ref{eq:55})
which in fact is the $U(1)\times SU(2)$ gauge structure of generalized
current associated with dyons consisting point like electromagnetic
$U(1)$ gauge structure having the four current $J_{\mu}$ followed
by $SU(2)$ like extended Yang - Mill's gauge structure with non Abelian
nature four current $J_{\mu}^{a}$. Equation (\ref{eq:56}) represents
the split octonion derivative for dyonic fields in non - symmetric
theory, which in fact is the double fold realization of quaternion
derivative. With the help of equation (\ref{eq:56}) we have obtained
equations (\ref{eq:57},\ref{eq:58}) and equation (\ref{eq:59}),
respectively defines the double fold $U(1)\times SU(2)$ gauge structures
of quaternion tangent space and generalized Dirac-Maxwell's equation
for dyonic fields. Also equation (\ref{eq:60}) represents the continuity
equation for dyonic fields in octonionic tangent space. It has been
shown that the theory of dynamics of electric and magnetic charges
is reproduced from the generalized theory of dyons using complex,
quaternion and octonion tangent spaces. Equation (\ref{eq:68}) illustrates
the covariant derivative for generalized fields of dyons in the complex
tangent space of Einstein - Schrödinger non - symmetric theory, where
$q^{\star}V_{\mu}$ is represented by equation (\ref{eq:69}). Consequently
equation (\ref{eq:70}) is the Dirac equation for generalized fields
of dyons in complex tangent space, which is invariant under gauge
transformation and Lorentz transformation as well. Equation (\ref{eq:71})
represents the covariant derivative in quaternionic tangent space.
In equation (\ref{eq:72}) $e$ and $g$ are electric and magnetic
charges of dyons and $\varepsilon$ and $\varepsilon'$ are Yang -
Mill's coupling constants associated with the isotopic spin coupling
parameters due to the presence of electric and magnetic charges respectively.
Thus equation (\ref{eq:73}) represents the Dirac equation for generalized
fields of dyons in quaternionic space of ES non - symmetric theory.
Similarly equations (\ref{eq:69}) represents the Dirac equation for
generalized fields of dyons in octonion tangent space if $D_{\mu}$
is described in its split octonion form given by equation (\ref{eq:74}).
Here we see that the Dirac equation in the octonion tangent space
is the doubly fold structure of quaternionic tangent space and is
described in terms of Zorn's vector matrix realization of split octonion
basis elements. As such, the fore going analysis describe the further
extension of ES non symmetric metrics successfully and consistently
in terms of three hyper complex number system namely complex, quaternion
and octonion without imposing extra constrains. So, in nutshell, the
present theory describes the combined gauge structures $GL(R)\otimes U(1)_{e}\otimes U(1)_{m}\otimes SU(2)_{e}\otimes SU(2)_{m}$
where $GL(R)$ describes Gravity, $U(1)_{e}$demonstrates the electromagnetism
due to the presence of electric charge, $U(1)_{m}$ is responsible
for the electromagnetism due to magnetic monopole, $SU(2)_{e}$demonstrates
the Yang Mill's field due to the presence of electric charge while
$SU(2)_{m}$gives rise the another Yang-Mills field due to the presence
of magnetic monopole. It has also been shown that this unified picture
reproduces the Gravity, electromagnetism and theory of Yang-Mill's
field in the absence of magnetic monopole. Accordingly we have obtained
the generalized Dirac equation for dyons from the covariant derivatives
in terms of complex, quaternionic and octonionic tangent spaces.

\textbf{Acknowledgment}- The work is supported by Uttarakhand Council
of Science and Technology, Dehradun. One of us OPSN is thankful to
Chinese Academy of Sciences and Third world Academy of Sciences for
awarding him CAS-TWAS visiting scholar fellowship to pursue a research
program in China. He is also grateful to Professor Tianjun Li for
his hospitality at Institute of Theoretical Physics, Beijing, China.

\end{document}